
\magnification=\magstephalf
\newbox\SlashedBox
\def\slashed#1{\setbox\SlashedBox=\hbox{#1}
\hbox to 0pt{\hbox to 1\wd\SlashedBox{\hfil/\hfil}\hss}#1}
\def\hboxtosizeof#1#2{\setbox\SlashedBox=\hbox{#1}
\hbox to 1\wd\SlashedBox{#2}}

\def\mathslashed#1{\setbox\SlashedBox=\hbox{$#1$}
\hbox to 0pt{\hbox to 1\wd\SlashedBox{\hfil/\hfil}\hss}#1}
\def\partialslash{\mathslashed{\partial}}

\def\ifsmall{\iffalse}  
\def\titlepagefont{}  

\def\DefineTeXgraphics{%
\special{ps::[global] /TeXgraphics { } def}}  

\def\today{\ifcase\month\or January\or February\or March\or April\or May
\or June\or July\or August\or September\or October\or November\or
December\fi\space\number\day, \number\year}
\def\eatPrefix19{}
\def\Year{\expandafter\eatPrefix\the\year}
\newcount\hours \newcount\minutes
\def\monthname{\ifcase\month\or
January\or February\or March\or April\or May\or June\or July\or
August\or September\or October\or November\or December\fi}
\def\shortmonthname{\ifcase\month\or
Jan\or Feb\or Mar\or Apr\or May\or Jun\or Jul\or
Aug\or Sep\or Oct\or Nov\or Dec\fi}

\def\TimeStamp{\hours\the\time\divide\hours by60%
\minutes -\the\time\divide\minutes by60\multiply\minutes by60%
\advance\minutes by\the\time%
${\rm \shortmonthname}\cdot\if\day<10{}0\fi\the\day\cdot\the\year%
\qquad\the\hours:\if\minutes<10{}0\fi\the\minutes$}




\def\Title#1{%
\vskip 1in{\titlefont\centerline{#1}}\vskip .5in}

\def\Date#1{\leftline{#1}\tenrm\supereject%
\global\hsize=\hsbody\global\hoffset=\hbodyoffset%
\footline={\hss\tenrm\folio\hss}}

\newif\ifdraftmode
\newif\ifleftlabels  

\def\nolabels{\def\wrlabeL##1{}\def\eqlabeL##1{}\def\reflabeL##1{}}
\def\writelabels{\def\wrlabeL##1{\leavevmode\vadjust{\rlap{\smash%
{\line{{\escapechar=` \hfill\rlap{\sevenrm\hskip.03in\string##1}}}}}}}%
\def\eqlabeL##1{{\escapechar-1\rlap{\sevenrm\hskip.05in\string##1}}}%
\def\reflabeL##1{\noexpand\rlap{\noexpand\sevenrm[\string##1]}}}
\def\writeleftlabels{\def\wrlabeL##1{\leavevmode\vadjust{\rlap{\smash%
{\line{{\escapechar=` \hfill\rlap{\sevenrm\hskip.03in\string##1}}}}}}}%
\def\eqlabeL##1{{\escapechar-1%
\rlap{\sixrm\hskip.05in\string##1}%
\llap{\sevenrm\string##1\hskip.03in\hbox to \hsize{}}}}%
\def\reflabeL##1{\noexpand\rlap{\noexpand\sevenrm[\string##1]}}}
\nolabels

\newdimen\fullhsize
\newdimen\hstitle
\hstitle=\hsize 
\newdimen\hsbody
\hsbody=\hsize 
\newdimen\hbodyoffset
\hbodyoffset=\hoffset 
\newbox\leftpage
\def\abstract#1{#1}
\def\rotated{\special{ps: landscape}
\magnification=1000  
\baselineskip=14pt
\global\hstitle=9truein\global\hsbody=4.75truein
\global\vsize=7truein\global\voffset=-.31truein
\global\hoffset=-0.54in\global\hbodyoffset=-.54truein
\global\fullhsize=10truein
\def\DefineTeXgraphics{%
\special{ps::[global]
/TeXgraphics {currentpoint translate 0.7 0.7 scale
              -80 0.72 mul -1000 0.72 mul translate} def}}
\let\lr=L
\def\ifsmall{\iftrue}
\def\titlepagefont{\twelvepoint}
\trueseventeenpoint
\def\almostshipout##1{\if L\lr \count1=1
      \global\setbox\leftpage=##1 \global\let\lr=R
   \else \count1=2
      \shipout\vbox{\hbox to\fullhsize{\box\leftpage\hfil##1}}
      \global\let\lr=L\fi}

\output={\ifnum\count0=1 
 \shipout\vbox{\hbox to \fullhsize{\hfill\pagebody\hfill}}\advancepageno
 \else
 \almostshipout{\leftline{\vbox{\pagebody\makefootline}}}\advancepageno
 \fi}

\def\abstract##1{{\leftskip=1.5in\rightskip=1.5in ##1\par}} }

\def\linemessage#1{\immediate\write16{#1}}

\global\newcount\secno \global\secno=0
\global\newcount\appno \global\appno=0
\global\newcount\meqno \global\meqno=1
\global\newcount\subsecno \global\subsecno=0
\global\newcount\figno \global\figno=0

\newif\ifAnyCounterChanged
\let\terminator=\relax
\def\normalize#1{\ifx#1\terminator\let\next=\relax\else%
\if#1i\aftergroup i\else\if#1v\aftergroup v\else\if#1x\aftergroup x%
\else\if#1l\aftergroup l\else\if#1c\aftergroup c\else%
\if#1m\aftergroup m\else%
\if#1I\aftergroup I\else\if#1V\aftergroup V\else\if#1X\aftergroup X%
\else\if#1L\aftergroup L\else\if#1C\aftergroup C\else%
\if#1M\aftergroup M\else\aftergroup#1\fi\fi\fi\fi\fi\fi\fi\fi\fi\fi\fi\fi%
\let\next=\normalize\fi%
\next}
\def\makeNormal#1#2{\def\doNormalDef{\edef#1}\begingroup%
\aftergroup\doNormalDef\aftergroup{\normalize#2\terminator\aftergroup}%
\endgroup}

\def\warnIfChanged#1#2{%
\ifundef#1
\else\begingroup%
\edef\oldDefinitionOfCounter{#1}\edef\newDefinitionOfCounter{#2}%
\ifx\oldDefinitionOfCounter\newDefinitionOfCounter%
\else%
\linemessage{Warning: definition of \noexpand#1 has changed.}%
\global\AnyCounterChangedtrue\fi\endgroup\fi}

\def\Section#1{\global\advance\secno by1\relax\global\meqno=1%
\global\subsecno=0%
\bigbreak\bigskip
\centerline{\twelvepoint \bf %
\the\secno. #1}%
\par\nobreak\medskip\nobreak}
\def\tagsection#1{%
\warnIfChanged#1{\the\secno}%
\xdef#1{\the\secno}%
\ifWritingAuxFile\immediate\write\auxfile{\noexpand\xdef\noexpand#1{#1}}\fi%
}
\def\section{\Section}
\def\Subsection#1{\global\advance\subsecno by1\relax\medskip %
\leftline{\bf\the\secno.\the\subsecno\ #1}%
\par\nobreak\smallskip\nobreak}
\def\tagsubsection#1{%
\warnIfChanged#1{\the\secno.\the\subsecno}%
\xdef#1{\the\secno.\the\subsecno}%
\ifWritingAuxFile\immediate\write\auxfile{\noexpand\xdef\noexpand#1{#1}}\fi%
}

\def\subsection{\Subsection}

\def\romappno{\uppercase\expandafter{\romannumeral\appno}}
\def\makeNormalizedRomappno{%
\expandafter\makeNormal\expandafter\normalizedromappno%
\expandafter{\romannumeral\appno}%
\edef\normalizedromappno{\uppercase{\normalizedromappno}}}
\def\Appendix#1{\global\advance\appno by1\relax\global\meqno=1\global\secno=0
\bigbreak\bigskip
\centerline{\twelvepoint \bf Appendix %
\romappno. #1}%
\par\nobreak\medskip\nobreak}
\def\tagappendix#1{\makeNormalizedRomappno%
\warnIfChanged#1{\normalizedromappno}%
\xdef#1{\normalizedromappno}%
\ifWritingAuxFile\immediate\write\auxfile{\noexpand\xdef\noexpand#1{#1}}\fi%
}
\def\appendix{\Appendix}

\def\eqn#1{\makeNormalizedRomappno%
\ifnum\secno>0%
  \warnIfChanged#1{\the\secno.\the\meqno}%
  \eqno(\the\secno.\the\meqno)\xdef#1{\the\secno.\the\meqno}%
     \global\advance\meqno by1
\else\ifnum\appno>0%
  \warnIfChanged#1{\normalizedromappno.\the\meqno}%
  \eqno({\rm\romappno}.\the\meqno)%
      \xdef#1{\normalizedromappno.\the\meqno}%
     \global\advance\meqno by1
\else%
  \warnIfChanged#1{\the\meqno}%
  \eqno(\the\meqno)\xdef#1{\the\meqno}%
     \global\advance\meqno by1
\fi\fi%
\eqlabeL#1%
\ifWritingAuxFile\immediate\write\auxfile{\noexpand\xdef\noexpand#1{#1}}\fi%
}
\def\defeqn#1{\makeNormalizedRomappno%
\ifnum\secno>0%
  \warnIfChanged#1{\the\secno.\the\meqno}%
  \xdef#1{\the\secno.\the\meqno}%
     \global\advance\meqno by1
\else\ifnum\appno>0%
  \warnIfChanged#1{\normalizedromappno.\the\meqno}%
  \xdef#1{\normalizedromappno.\the\meqno}%
     \global\advance\meqno by1
\else%
  \warnIfChanged#1{\the\meqno}%
  \xdef#1{\the\meqno}%
     \global\advance\meqno by1
\fi\fi%
\eqlabeL#1%
\ifWritingAuxFile\immediate\write\auxfile{\noexpand\xdef\noexpand#1{#1}}\fi%
}
\def\anoneqn{\makeNormalizedRomappno%
\ifnum\secno>0
  \eqno(\the\secno.\the\meqno)%
     \global\advance\meqno by1
\else\ifnum\appno>0
  \eqno({\rm\normalizedromappno}.\the\meqno)%
     \global\advance\meqno by1
\else
  \eqno(\the\meqno)%
     \global\advance\meqno by1
\fi\fi%
}
\def\mfig#1#2{\global\advance\figno by1%
\relax#1\the\figno%
\warnIfChanged#2{\the\figno}%
\edef#2{\the\figno}%
\reflabeL#2%
\ifWritingAuxFile\immediate\write\auxfile{\noexpand\xdef\noexpand#2{#2}}\fi%
}

\catcode`@=11 

\font\ninerm=cmr9
\font\eightrm=cmr8
\font\sixrm=cmr6

\def\loadtrueseventeenpoint{
 \font\seventeenrm=cmr10 at 17.28truept
 \font\seventeeni=cmmi10 at 17.28truept
 \font\seventeenbf=cmbx10 at 17.28truept
 \font\seventeenit=cmti10 at 17.28truept
 \font\seventeensl=cmsl10 at 17.28truept
 \font\seventeensy=cmsy10 at 17.28truept
}
\def\loadfourteenpoint{
\font\fourteenrm=cmr10 at 14.4pt
\font\fourteeni=cmmi10 at 14.4pt
\font\fourteenit=cmti10 at 14.4pt
\font\fourteensl=cmsl10 at 14.4pt
\font\fourteensy=cmsy10 at 14.4pt
\font\fourteenbf=cmbx10 at 14.4pt
}
\def\loadtruetwelvepoint{
\font\twelverm=cmr10 at 12truept
\font\twelvei=cmmi10 at 12truept
\font\twelveit=cmti10 at 12truept
\font\twelvesl=cmsl10 at 12truept
\font\twelvesy=cmsy10 at 12truept
\font\twelvebf=cmbx10 at 12truept
}

\font\ninei=cmmi9
\font\eighti=cmmi8
\font\sixi=cmmi6
\skewchar\ninei='177 \skewchar\eighti='177 \skewchar\sixi='177

\font\ninesy=cmsy9
\font\eightsy=cmsy8
\font\sixsy=cmsy6
\skewchar\ninesy='60 \skewchar\eightsy='60 \skewchar\sixsy='60

\font\ninebf=cmbx9
\font\eightbf=cmbx8
\font\sixbf=cmbx6

\font\ninett=cmtt9
\font\eighttt=cmtt8

\hyphenchar\tentt=-1 
\hyphenchar\ninett=-1
\hyphenchar\eighttt=-1

\font\ninesl=cmsl9
\font\eightsl=cmsl8

\font\nineit=cmti9
\font\eightit=cmti8


\newskip\ttglue
\def\tenpoint{\def\rm{\fam0\tenrm}%
  \textfont0=\tenrm \scriptfont0=\sevenrm \scriptscriptfont0=\fiverm
  \textfont1=\teni \scriptfont1=\seveni \scriptscriptfont1=\fivei
  \textfont2=\tensy \scriptfont2=\sevensy \scriptscriptfont2=\fivesy
  \textfont3=\tenex \scriptfont3=\tenex \scriptscriptfont3=\tenex
  \def\it{\fam\itfam\tenit}\textfont\itfam=\tenit
  \def\sl{\fam\slfam\tensl}\textfont\slfam=\tensl
  \def\bf{\fam\bffam\tenbf}\textfont\bffam=\tenbf \scriptfont\bffam=\sevenbf
  \scriptscriptfont\bffam=\fivebf
  \normalbaselineskip=12pt
  \let\sc=\eightrm
  \let\big=\tenbig
  \setbox\strutbox=\hbox{\vrule height8.5pt depth3.5pt width\z@}%
  \normalbaselines\rm}

\def\twelvepoint{\def\rm{\fam0\twelverm}%
  \textfont0=\twelverm \scriptfont0=\ninerm \scriptscriptfont0=\sevenrm
  \textfont1=\twelvei \scriptfont1=\ninei \scriptscriptfont1=\seveni
  \textfont2=\twelvesy \scriptfont2=\ninesy \scriptscriptfont2=\sevensy
  \textfont3=\tenex \scriptfont3=\tenex \scriptscriptfont3=\tenex
  \def\it{\fam\itfam\twelveit}\textfont\itfam=\twelveit
  \def\sl{\fam\slfam\twelvesl}\textfont\slfam=\twelvesl
  \def\bf{\fam\bffam\twelvebf}\textfont\bffam=\twelvebf
\scriptfont\bffam=\ninebf
  \scriptscriptfont\bffam=\sevenbf
  \normalbaselineskip=12pt
  \let\sc=\eightrm
  \let\big=\tenbig
  \setbox\strutbox=\hbox{\vrule height8.5pt depth3.5pt width\z@}%
  \normalbaselines\rm}

\def\fourteenpoint{\def\rm{\fam0\fourteenrm}%
  \textfont0=\fourteenrm \scriptfont0=\tenrm \scriptscriptfont0=\sevenrm
  \textfont1=\fourteeni \scriptfont1=\teni \scriptscriptfont1=\seveni
  \textfont2=\fourteensy \scriptfont2=\tensy \scriptscriptfont2=\sevensy
  \textfont3=\tenex \scriptfont3=\tenex \scriptscriptfont3=\tenex
  \def\it{\fam\itfam\fourteenit}\textfont\itfam=\fourteenit
  \def\sl{\fam\slfam\fourteensl}\textfont\slfam=\fourteensl
  \def\bf{\fam\bffam\fourteenbf}\textfont\bffam=\fourteenbf%
  \scriptfont\bffam=\tenbf
  \scriptscriptfont\bffam=\sevenbf
  \normalbaselineskip=17pt
  \let\sc=\elevenrm
  \let\big=\tenbig
  \setbox\strutbox=\hbox{\vrule height8.5pt depth3.5pt width\z@}%
  \normalbaselines\rm}

\def\seventeenpoint{\def\rm{\fam0\seventeenrm}%
  \textfont0=\seventeenrm \scriptfont0=\fourteenrm \scriptscriptfont0=\tenrm
  \textfont1=\seventeeni \scriptfont1=\fourteeni \scriptscriptfont1=\teni
  \textfont2=\seventeensy \scriptfont2=\fourteensy \scriptscriptfont2=\tensy
  \textfont3=\tenex \scriptfont3=\tenex \scriptscriptfont3=\tenex
  \def\it{\fam\itfam\seventeenit}\textfont\itfam=\seventeenit
  \def\sl{\fam\slfam\seventeensl}\textfont\slfam=\seventeensl
  \def\bf{\fam\bffam\seventeenbf}\textfont\bffam=\seventeenbf%
  \scriptfont\bffam=\fourteenbf
  \scriptscriptfont\bffam=\twelvebf
  \normalbaselineskip=21pt
  \let\sc=\fourteenrm
  \let\big=\tenbig
  \setbox\strutbox=\hbox{\vrule height 12pt depth 6pt width\z@}%
  \normalbaselines\rm}

\def\ninepoint{\def\rm{\fam0\ninerm}%
  \textfont0=\ninerm \scriptfont0=\sixrm \scriptscriptfont0=\fiverm
  \textfont1=\ninei \scriptfont1=\sixi \scriptscriptfont1=\fivei
  \textfont2=\ninesy \scriptfont2=\sixsy \scriptscriptfont2=\fivesy
  \textfont3=\tenex \scriptfont3=\tenex \scriptscriptfont3=\tenex
  \def\it{\fam\itfam\nineit}\textfont\itfam=\nineit
  \def\sl{\fam\slfam\ninesl}\textfont\slfam=\ninesl
  \def\bf{\fam\bffam\ninebf}\textfont\bffam=\ninebf \scriptfont\bffam=\sixbf
  \scriptscriptfont\bffam=\fivebf
  \normalbaselineskip=11pt
  \let\sc=\sevenrm
  \let\big=\ninebig
  \setbox\strutbox=\hbox{\vrule height8pt depth3pt width\z@}%
  \normalbaselines\rm}

\def\eightpoint{\def\rm{\fam0\eightrm}%
  \textfont0=\eightrm \scriptfont0=\sixrm \scriptscriptfont0=\fiverm%
  \textfont1=\eighti \scriptfont1=\sixi \scriptscriptfont1=\fivei%
  \textfont2=\eightsy \scriptfont2=\sixsy \scriptscriptfont2=\fivesy%
  \textfont3=\tenex \scriptfont3=\tenex \scriptscriptfont3=\tenex%
  \def\it{\fam\itfam\eightit}\textfont\itfam=\eightit%
  \def\sl{\fam\slfam\eightsl}\textfont\slfam=\eightsl%
  \def\bf{\fam\bffam\eightbf}\textfont\bffam=\eightbf \scriptfont\bffam=\sixbf%
  \scriptscriptfont\bffam=\fivebf%
  \normalbaselineskip=9pt%
  \let\sc=\sixrm%
  \let\big=\eightbig%
  \setbox\strutbox=\hbox{\vrule height7pt depth2pt width\z@}%
  \normalbaselines\rm}

\def\tenbig#1{{\hbox{$\left#1\vbox to8.5pt{}\right.\n@space$}}}
\def\ninebig#1{{\hbox{$\textfont0=\tenrm\textfont2=\tensy
  \left#1\vbox to7.25pt{}\right.\n@space$}}}
\def\eightbig#1{{\hbox{$\textfont0=\ninerm\textfont2=\ninesy
  \left#1\vbox to6.5pt{}\right.\n@space$}}}

\def\footnote#1{\edef\@sf{\spacefactor\the\spacefactor}#1\@sf
      \insert\footins\bgroup\eightpoint
      \interlinepenalty100 \let\par=\endgraf
        \leftskip=\z@skip \rightskip=\z@skip
        \splittopskip=10pt plus 1pt minus 1pt \floatingpenalty=20000
        \smallskip\item{#1}\bgroup\strut\aftergroup\@foot\let\next}
\skip\footins=12pt plus 2pt minus 4pt 
\dimen\footins=30pc 

\newinsert\margin
\dimen\margin=\maxdimen
\def\titlefont{\seventeenpoint}
\loadtruetwelvepoint 
\loadtrueseventeenpoint
\catcode`\@=\active
\catcode`@=12  
\catcode`\"=\active

\def\eatOne#1{}
\def\ifundef#1{\expandafter\ifx%
\csname\expandafter\eatOne\string#1\endcsname\relax}
\def\notTrue{\iffalse}\def\isTrue{\iftrue}
\def\ifdef#1{{\ifundef#1%
\aftergroup\notTrue\else\aftergroup\isTrue\fi}}
\def\use#1{\ifundef#1\linemessage{Warning: \string#1 is undefined.}%
{\tt \string#1}\else#1\fi}


\global\newcount\refno \global\refno=1
\newwrite\rfile
\newlinechar=`\^^J
\def\ref#1#2{\the\refno\nref#1{#2}}
\def\nref#1#2{\xdef#1{\the\refno}%
\ifnum\refno=1\immediate\openout\rfile=refs.tmp\fi%
\immediate\write\rfile{\noexpand\item{[\noexpand#1]\ }#2.}%
\global\advance\refno by1}
\def\lref#1#2{\the\refno\xdef#1{\the\refno}%
\ifnum\refno=1\immediate\openout\rfile=refs.tmp\fi%
\immediate\write\rfile{\noexpand\item{[\noexpand#1]\ }#2\semi}%
\global\advance\refno by1}
\def\cref#1{\immediate\write\rfile{#1\semi}}

\def\semi{;\hfil\noexpand\break}

\def\inputAuxIfPresent#1{\immediate\openin1=#1
\ifeof1\message{No file \auxfileName; I'll create one.
}\else\closein1\relax\input\auxfileName\fi%
}
\def\NPB{Nucl.\ Phys.\ B}
\def\PRL{Phys.\ Rev.\ Lett.\ }
\def\PRD{Phys.\ Rev.\ D}
\def\PLB{Phys.\ Lett.\ B}

\newif\ifWritingAuxFile
\newwrite\auxfile
\def\SetUpAuxFile{%
\xdef\auxfileName{\jobname.aux}%
\inputAuxIfPresent{\auxfileName}%
\WritingAuxFiletrue%
\immediate\openout\auxfile=\auxfileName}

\def\L{\left(}\def\R{\right)}

\def\LB{\left[}\def\RB{\right]}

\def\bye{\par\vfill\supereject%
\ifAnyCounterChanged\linemessage{
Some counters have changed.  Re-run tex to fix them up.}\fi%
\end}

\input epsf 
\SetUpAuxFile
\loadfourteenpoint

\overfullrule 0pt
\hfuzz 20pt

\def\Cutoff{\Lambda}
\def\XSB{\Lambda_\chi}
\def\Strong{\Lambda_s}
\def\Lag{{\cal L}}
\def\psibar{\overline{\psi}}
\def\Ord{{\cal O}}
\def\Dslash{\slashed{D}}
\def\lappeq{\mathrel{ \rlap{\raise.5ex\hbox{$<$}}
                      {\lower.5ex\hbox{$\sim$}}  } }
\def\Op#1{O_{#1}}

\iftrue
\nopagenumbers
\rightline{CERN-TH.6767/92}
\rightline{hep-ph/9302289}

\leftlabelstrue
\Title{Light Composite Vector Bosons}

\centerline{David A. Kosower${}^\dagger$}
\smallskip
\centerline{\it Theory Division}
\centerline{\it CERN}
\centerline{\it CH-1211 Geneva 23}
\centerline{\it Switzerland}
\centerline{\tt kosower@dxcern.cern.ch}

\vskip 0.2in\baselineskip13truept

\vskip 0.75truein
\centerline{\bf Abstract}

{\narrower

In gauge theories with slowly-running coupling constants,
it may be possible for four-fermion
operators to be nearly marginal.  Such operators
can possess asymptotically weak
couplings, and can plausibly give rise to light
composite vector mesons.
}

\baselineskip17pt

\vfill\vskip 0.2in
\noindent\hrule width 3.6in\hfil\break
${}^{\dagger}$ On leave from the Centre d'Etudes de Saclay,
F-91191 Gif-sur-Yvette cedex, France\hfil\break
\vskip 0.3truein
\leftline{CERN-TH.6767/92}
\leftline{December, 1992}
\Date{}

\line{}
\fi

\baselineskip17pt

The so-called Standard Model of accessible-energy particle physics, in
spite of its increasingly precise experimental confirmation, remains
unsatisfactory for several theoretical reasons.  One of these is the
hierarchy problem: it is unnatural for the Higgs scalar to appear in
the theory at all, since its mass is not protected by any symmetry.
Another is the replication of flavors and associated mass and
mixing-angle parameters: these may be technically natural, but are
puzzling nonetheless.

These theoretical defects have prompted the development of many
extensions to the Standard Model over the past two decades.  Sensible
extensions 
should possess a limit in which they reduce to the Standard Model, as
this allows them to explain why the latter is such a
successful effective theory at currently accessible energies.  Such
extensions can solve the
hierarchy problem: both supersymmetry and
technicolor~[\ref\Technicolor{S. Weinberg, \PRD 13:974 (1976); \PRD 19:1277
(1979)\semi
L. Susskind, \PRD 20:2619 (1979)}]
models (with or without grand unification) do.  The
former protects a fine-tuning, while the latter also explains the
origin of a small scale (compared to the other known scale, the Planck
scale).  The flavor problem, on the other hand, is evidently quite
hard: supersymmetric models do not even
address it, while the literature is littered with the corpses of
extensions to technicolor models that have succumbed either to the
presence of excessive flavor-changing neutral currents or to the
inadmissibility of a heavy top-quark, or to both.


Both of these approaches presume that the flavor physics underlying the
Standard Model can be explained purely in terms of weak-coupling
physics, possibly supplemented by a scaled-up version of QCD.  This
is not necessarily wrong, but does seem a bit presumptuous, especially
in light of the growing evidence that one of the parameters
(the top Yukawa coupling) may be large enough
to produce non-trivial dynamics.  It thus seems sensible to search
for other extensions to the Standard Model in which strong-coupling
dynamics plays a role.

One step in this direction was taken by Bardeen, Hill, and Lindner
(BHL)~[\ref\BHL{W. A. Bardeen, C. T. Hill, and M. Lindner,
\PRD 41:1647 (1990)}],
following earlier work~[\ref\Nambu{Y. Nambu, Univ.\ of Chicago
preprint EFI 89--08 (1989)\semi
V. A. Miransky, M. Tanabashi, and K. Yamawaki, \PLB 221:177 (1989),
Mod.\ Phys.\ Lett.\ A4:1043 (1989)}].  They
introduced a Nambu--Jona-Lasinio--type~[\ref\NJL{Y. Nambu and
G. Jona-Lasinio, Phys.\ Rev.\ 122:345 (1961)}]
four-fermion coupling of the top quark with a large dynamically-generated
anomalous dimension; at lower energies, it
becomes strong, generating an $SU(2)\times U(1)$-breaking condensate,
along with a composite Higgs scalar.  In the limit originally considered
by these authors, their model was later shown to be completely
equivalent to the Standard
Model~[\ref\HHJKS{%
A. Hasenfratz, P. Hasenfratz, K. Jansen, J. Kuti, and Y. Shen,
  \NPB 365:79 (1991)\semi
J. Zinn-Justin, \NPB 367:105 (1991)\semi
J. Soto, \PLB 280:75 (1992)}].  Far from being a defect, this is
in fact a virtue, as explained above.  Interesting extensions
may be obtained not by taking the BHL `cutoff' (in their
language) very high, but rather in taking it as {\it low\/} as possible
without violating known experimental constraints [\ref\NewBHL{%
C. T. Hill, \PLB 266:419 (1991)\semi
M. Lindner and D. Ross, \NPB 370:30 (1992)\semi
H. M. Chesterman and S. F. King, \PRD 45:297 (1992)\semi
S. P. Martin, \PRD 45:4283 (1992)}].

  In this letter, I explore another possible avenue for generating
extensions to the Standard Model: rather than rewriting the Yukawa
interaction in terms of four-fermion couplings, and generating a
composite scalar, I rewrite the spontaneously-broken
gauge interactions in terms
of four-fermion operators, generating a light, composite, vector field.
This idea is in fact nearly as old as I
am~[\ref\BJ{J. D. Bjorken, Ann.\  Phys.\ 24:174 (1963)},%
\ref\Early{I. Bialynicki-Birula, Phys.\ Rev.\ 130:465 (1963)\semi
G. S. Guralnik, Phys.\ Rev.\ 136B:1404 (1964)\semi
T. Eguchi and H. Sugawara, \PRD10:4257 (1974)\semi
C. Bender, F. Cooper, and G. Guralnik, Ann.\ Phys.\ 109:165 (1977)},%
\ref\Late{T. Banks and A. Zaks, \NPB 184:303 (1981)\semi
M. Veltman, Acta Phys.\ Pol.\ B12:437 (1981)\semi
M. Peskin, in Proceedings of the 1981 Int'l Symposium on Lepton and Photon
Interactions (Bonn, 1981; ed.\ W. Pfeil)%
},%
\ref\EffectiveAction{%
T. Eguchi, \PRD14:2755 (1976)\semi
M. Suzuki, \PRD37:210 (1988)\semi
A. Cohen, H. Georgi, and E. H. Simmons, \PRD38:405 (1988)\semi
A. Hasenfratz and P. Hasenfratz, preprint BUTP-92/28 (1992)}].
The sensible attempts have concluded that it is both necessary
(for consistency [\ref\Consistency{%
T. D. Lee and B. Zumino, Phys.\ Rev.\ 163:1667 (1967)\semi
J. M. Cornwall, D. N. Levin, and G. Tiktopoulos, \PRL 30:1268 (1973);
\PRD 10:1145 (1974)}])
and natural for the couplings of the light vector to approach
the form of gauge couplings at low energies.
For reasons I will review below, however, these
attempts have confined their attention to the low-energy effective
theory beneath the binding scale, and have not been able to explore
the possible origins of such a theory at high energies.
What is new below is the
observation that gauge theories with slowly-running couplings ---
so-called walking gauge theories~[\ref\Holdom{B. Holdom,
\PLB 150:301 (1985)},%
\ref\WTC{T. Akiba and T. Yanagida, \PLB 169:432 (1986)\semi
T. Appelquist, D. Karabali, and L. C. R. Wijewardhana, \PRL 57:957 (1986)\semi
T. Appelquist and L. C. R. Wijewardhana, \PRD 35:774 (1987)\semi
M. Bando, T. Morozumi, H. So, and K. Yamawaki, \PRL 59:389 (1987)},%
\ref\WalkingSample{T. Appelquist
and L. C. R. Wijewardhana, \PRD 36:568 (1987)}]
--- allow us to do so.

Let me begin by paraphrasing previous analyses of a theory of
$N$ massive fermions with an attractive vector interaction,
$$
\Lag = \psibar (\partialslash + M) \psi + {f\over \Cutoff^2}
  \L \psibar\gamma_\mu T^a\psi\R^2\;,
\eqn\BasicL
$$
where $T^a$ are the generators of $SU(N)$ in the fundamental
representation, and where $\Cutoff$
is a large scale ($\gg M$) where the four-fermion operator was
originally induced.  We may imagine evolving the lagrangian
down to smaller and smaller scales; when we cross $M$, we should
integrate out the massive fermions, since they will no longer be
present in the theory at lower energies.  As the various authors of
ref.~[\use\EffectiveAction]
show, when we do this, we will generate the action for an $SU(N)$
gauge field, with mass $\Cutoff/\sqrt{f}$, along with various
higher dimension operators which will be small for $p\ll M$.

Of course, we will only get to such momenta if the mass of the
vector is also much less than $M$; this in turn requires a
large $f$,
$$
f \gg \L{\Cutoff\over M}\R^2\;.
\eqn\fValue
$$
Were this reasonable, the presence of a light vector would be
understandable: the strong attractive interaction in the vector
channel binds the fermions deeply, lowering the mass of
the vector two-fermion state far below its naive threshold value of $2M$.

Unfortunately, from an effective field-theory point of view,
equation~(\fValue) is anything but reasonable, and it is certainly
not possible to interpret the theory between the scales $M$
and $\Cutoff$ as having the usual sort of weak-coupling
effective Lagrangian we are used to.

As we will see, the picture can change considerably if we take
the fermions to transform under a non-Abelian gauge interaction
with a `walking' coupling of sufficiently large value.  The basic
point is that such a walking gauge interaction (WGI for short)
can induce large anomalous dimensions for four-fermion operators,
perhaps sufficiently large to make operators like
the one in the Lagrangian~(\BasicL) nearly marginal (in the language of
the renormalization group).  (A related point was noted
by Bardeen, Leung, and
Love~[\ref\BLL{W. A. Bardeen, C. N. Leung, and S. T. Love, \PRL 56:1230\semi
C. N. Leung, S. T. Love, and W. A. Bardeen, \NPB 273:649 (1986)}]
in the context of quenched planar QED.)
This alone would be sufficient to
tame equation~(\fValue) to something like $f \gg 1$, but it is
not quite the end of the story.  For nearly marginal operators,
it is important to consider self-renormalization effects, and
as we shall see, an attractive operator like the one
in the Lagrangian~(\BasicL) is
in fact asymptotically weak,
so that $f(\Cutoff)$ can be perturbatively
small, yet nonetheless ultimately give rise to a light
vector state.

\def\Wf{\Psi}\def\Wfbar{\overline{\Wf}}
To understand this concretely, consider the following Lagrangian,
$$
\Lag = \Wfbar\Dslash \Wf + {f\over \Cutoff^2}
\L\Wfbar  T^a\gamma_\mu P_L\Wf\R^2 + \cdots
\eqn\WGIL$$
with $l$ flavors of Dirac fermions $\Wf$ transforming as an
$N_w$
of the walking gauge theory $SU(N_w)$, where $T^a$ are
the generators of the fundamental representation of $SU(l)$,
and where $P_L = (1-\gamma_5)/2$ is a left-chiral projection
operator.  (The dots indicate other fermions which do not
participate in the four-fermion interactions, but are required
in order to make the coupling walk.)
In an $SU(N)$ theory with $n_f$ fundamental Dirac fermions, the theory has
a running coupling with the two-loop beta
function~[\ref\TwoLoopBeta{D. R. T. Jones, \NPB 75:531 (1974)\semi
   W. E. Caswell, \PRL 33:244 (1974)}],
$$
\beta(\alpha) = -{1\over 2\pi} \L {11\over 3} N - {2\over3} n_f\R \alpha^2
     -{1\over8\pi^2}\L {34\over3} N^2 -{10\over3} N n_f-2 C_f n_f\R\alpha^3\;.
\anoneqn
$$
In a walking theory, the matter content is such that
the leading coefficient is much smaller than its canonical value
(say with $n_f=0$), while
the second coefficient is smaller yet in magnitude.
As a result, the coupling runs very slowly in the weak-coupling regime.
For a theory to walk at interesting values of the coupling constant,
we should require\footnote{${}^\dagger$}
  {The beta function in fact becomes renormalization-scheme
   dependent beyond two-loop
   order, so it is not clear what this statement means.  Anomalous
   dimensions of physical quantities calculated to all orders
   in perturbation theory will not suffer from this problem. }
 $\beta(\alpha)/\alpha \ll 1$
even at finite couplings of order the critical coupling for chiral-symmetry
breaking.
Typical walking theories have $\beta(\alpha)/\alpha \sim -0.1$ to $-0.3$ for
near-critical $\alpha$'s~[\use\WalkingSample] in contrast to
three-flavor QCD, which has $\beta(\alpha)/\alpha \sim -1.6$ for such
couplings.
Actually, we may choose to define this critical coupling $\alpha_c$ in
two ways: either as the coupling at which chiral-symmetry breaking
is triggered, or else the coupling at which the anomalous dimension for the
fermion self-energy becomes exactly $-1$.  (Cohen and
Georgi~[\ref\CG{A. Cohen and H. Georgi, \NPB 314:7 (1989)}]
argue that these two definitions are equivalent even beyond perturbation
theory.)
Within the resummed-rainbow approximation,
$$
\alpha_c = {\pi\over 3 C_F(N_w)}\;,
\anoneqn$$
where $C_F$ is the quadratic Casimir of the fundamental representation.

What happens to the four-fermion operator in the Lagrangian~(\WGIL)
under the influence of the gauge interactions?  For one thing, it
will mix with other operators, so we should consider the full set
of operators which mix under renormalization-group flow.  At first order
in perturbation theory (and thus within the analog of the
resummed-rainbow approximation), there are two such operators we
must consider,
$$
\Op\pm = {1\over2\Lambda^2}\LB\Wfbar^x T^a\gamma_\mu P_L\Wf_x\,
           \Wfbar^y T^a\gamma^\mu P_L\Wf_y
          \pm\Wfbar^x T^a\gamma_\mu P_L\Wf_y\,
           \Wfbar^y T^a\gamma^\mu P_L\Wf_x\RB\;,
\eqn\Operators$$
where $x,y$ denote WGI indices.
The four-fermion operator in equation~(\WGIL) is simply the sum of
$\Op+$ and $\Op-$.
(I have ignored the WGI penguin operator
$\Wfbar \tau^x_w\gamma_\mu P_L\Wf f_{xyz} D^y_\nu F^{z\mu\nu}$
 for reasons which will be explained
later, and possible non-perturbative effects such as instanton-induced
mixing for simplicity.)  The contributions of these operators
to Green functions will have the form
$$
\L1 - \gamma_0 \ln(\Lambda/p) + \cdots\R\,O\;,
\anoneqn$$
which at finite constant coupling we might rewrite in the form
$$
\L {p\over\Lambda}\R^{\gamma_0} O + \cdots
\anoneqn$$
In particular, if the perturbative
anomalous dimension~[\ref\WeakFF{%
G. Altarelli and L. Maiani, \PLB 52:351 (1974)\semi
F. J. Gilman and M. B. Wise, \PRD 20:2392 (1979)}]
$\gamma_0$ is negative, the operator will be enhanced
at scale $p$ compared to its naive strength of $(p/\Lambda)^2$.  Can
$|\gamma_0|$ be large enough to overcome this intrinsic suppression?

\topinsert\hfuzz 40pt
\noindent\hbox to 6truein{
\epsfxsize=6.5truein
\noindent\epsfbox{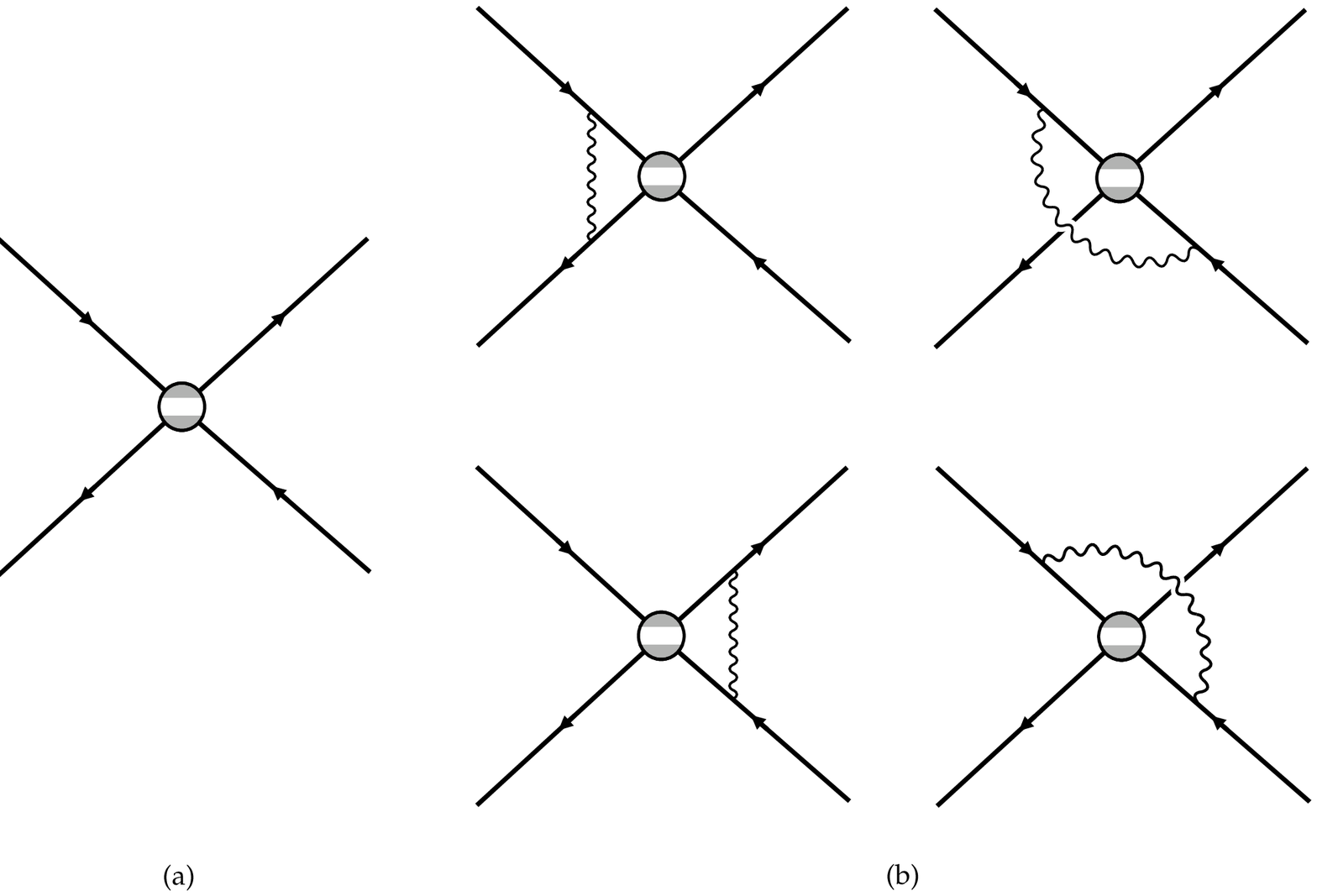}}
\hfil\break
\hbox to 6.5truein{
\hfil \vtop{
\noindent\eightpoint
Figure~1 (a) The left-hand side of the Schwinger-Dyson equation for the
four-fermion operator (b) The contributions to the right-hand side of
the Schwinger-Dyson equation in the Landau-gauge rainbow approximation.
                  \smallskip}\hfil}
\endinsert\hfuzz 20pt 

To go beyond weak coupling, we should
in principle, solve the (matrix) Schwinger-Dyson
equations generated (in Landau gauge)
by the collection of diagrams of fig.~1.
This is a rather difficult task.  To get an idea of the
qualitative behavior we might expect, however, we can proceed
as follows.  First, assume that the important mixing
effects are those given by the perturbative anomalous dimension matrix;
and that a multiplicatively-renormalized
operator $O_m$ scales simply according
to the momentum flowing in the vector channel.
Also, even though the fermion self-energy is enhanced compared
to its value in an ordinary gauge theory, it is still small
and hence can be neglected.  This suggests the
following simplification of the Schwinger-Dyson equation,
$$
O_m(s) = -{1\over\Omega_3}
    \int d^4k\; \gamma_0(\alpha(k)) {O_m(k^2)\over (k-p_1)^2 (k-p_2)^2}\;,
\anoneqn$$
where $\Omega_3=2\pi^2$ is the volume of the unit three-sphere.
Even this reduced form is hard to understand analytically; but we may
expect that at small momenta, the denominator will go as $\sim s^2$,
where $s$ is a typical invariant mass,
while at high momenta the invariant mass will be irrelevant, and
the denominator will go as $\sim k^4$.  We may also work in
the `standing' approximation in which the coupling does not
run at all.
Following this reasoning,
we arrive at the following caricature of the original equation,
$$
O_m(s) = -\gamma_0
    \int_0^{\sqrt{s}} k^3 dk\; {O_m(k^2)\over s^2}
    +\int_{\sqrt{s}}^\Lambda {dk\over k}\; O_m(k^2)\;.
\eqn\Caricature$$
If we now plug in an ansatz for the form of $O_m(s)\sim s^{\gamma/2}$,
perform the integrals, and extract the pieces proportional to $s^{\gamma/2}$,
we can derive the algebraic equation,
$$
1 = {4\gamma_0\over \gamma (4+\gamma)}\;.
\eqn\AnomalousDimension$$
For $\gamma_0$ small, this just yields the perturbative result,
$\gamma=\gamma_0$ (we discard the other solution); in general, we find
$$
\gamma = -2 + 2\sqrt{1+\gamma_0}\;.
\eqn\AD$$
As $\gamma_0\rightarrow -1$, the anomalous
dimension becomes large enough that the four-fermion operator, of
naive dimension six, can become nearly marginal: an anomalous dimension
of $-2+\delta$ implies an effective scaling dimension of $4+\delta$,
i.e. that of a marginally irrelevant operator.  Furthermore, the
anomalous dimension is {\it larger\/} in magnitude than would be
indicated by a linear extrapolation of perturbation theory, a
feature that is also true of the self-energy anomalous dimension
within the rainbow approximation.
(We should actually take the argument of $O_m$
in the first integral in equation~(\use\AnomalousDimension)
to be something like $k^2+s$; this does not
change the qualitative statements made here, though the numerical values of
the anomalous dimension $\gamma$ become somewhat smaller.)

Unfortunately, we also hit our first obstacle here.  If we rewrite
the perturbative coefficient $\gamma_0$ in terms of $\alpha/\alpha_c$,
we find
$$
\gamma_0 = - {r\over 2} \L{\alpha\over\alpha_c}\R\;,
\eqn\ratio
$$
where $r$ is a group-theoretic coefficient.  If we consider a more
general fermion representation, with $\Wf_{1,2}$ transforming in
representations $R_{1,2}$ of the WGI, and four-fermion operators
$$
O = {1\over\Lambda^2} M_{xyz} M_{\hat x\hat y z}
           \Wfbar_1^x T^a \gamma_\mu P_L\Wf_1^{\hat x}\,
           \Wfbar_2^y T^a \gamma_\mu P_L\Wf_2^{\hat y}\;,
\anoneqn$$
where the matrices $M$ are the Clebsch coefficients for
$R_1\otimes R_2 \rightarrow R_3$, and $x,y,z,\ldots$ represent sets of
WGI indices, then
$$
r = {C_2(R_1)+C_2(R_2)-C_2(R_3)\over \max\L C_2(R_1),C_2(R_2)\R}\;.
\anoneqn$$
($C_2(R)$ is the quadratic Casimir of representation $R$.)
If $R_1=\overline{R}_2$, and we take at $R_3=1$,
then $r=2$; but in this case
(which corresponds to a mixed left-right operator in QCD),
the relevant four-fermion operator presumably contributes
in the appropriate color-singlet {\it scalar\/} channel (and
also contributes to destabilizing the chirally-symmetric vacuum).
For other cases, $R_3$ cannot be the identity, and $r < 2$.
In particular,
if we stick to fundamental complex representations in vector-like
theories, then in fact
we find $r\leq 1$ for
$SU(N)$.
More generally, the best one seems to be able to do with
complex representations while
maintaining asymptotic freedom is with third-rank antisymmetric
tensor representations of $SU(7)$, for which $r=3/2$.
If we take the form of equation~(\use\AD) seriously,
the fact that $r<2$ implies that the coupling
will have grown large enough to trigger chiral-symmetry breaking
before it has gotten large enough to make any $LL$ four-fermion
operator nearly marginal.  Of course, in the full theory, the
form of this equation should not be taken too seriously; and it
may well be that the theory enhances the four-fermion anomalous dimensions
more at strong coupling than the corresponding one for the
fermion self-energy.  In this case, a four-fermion operator
could become nearly marginal at couplings smaller than that needed
to trigger chiral-symmetry breaking.  Even without hoping for
such a miracle, however, there are other avenues of escape from
this impasse.  I will mention two of them below.

\topinsert\hfuzz 40pt
\null\vskip 0.5truein
\noindent\hbox to 6truein{
\epsfxsize=5.5truein
\noindent\hskip0.5truein\epsfbox{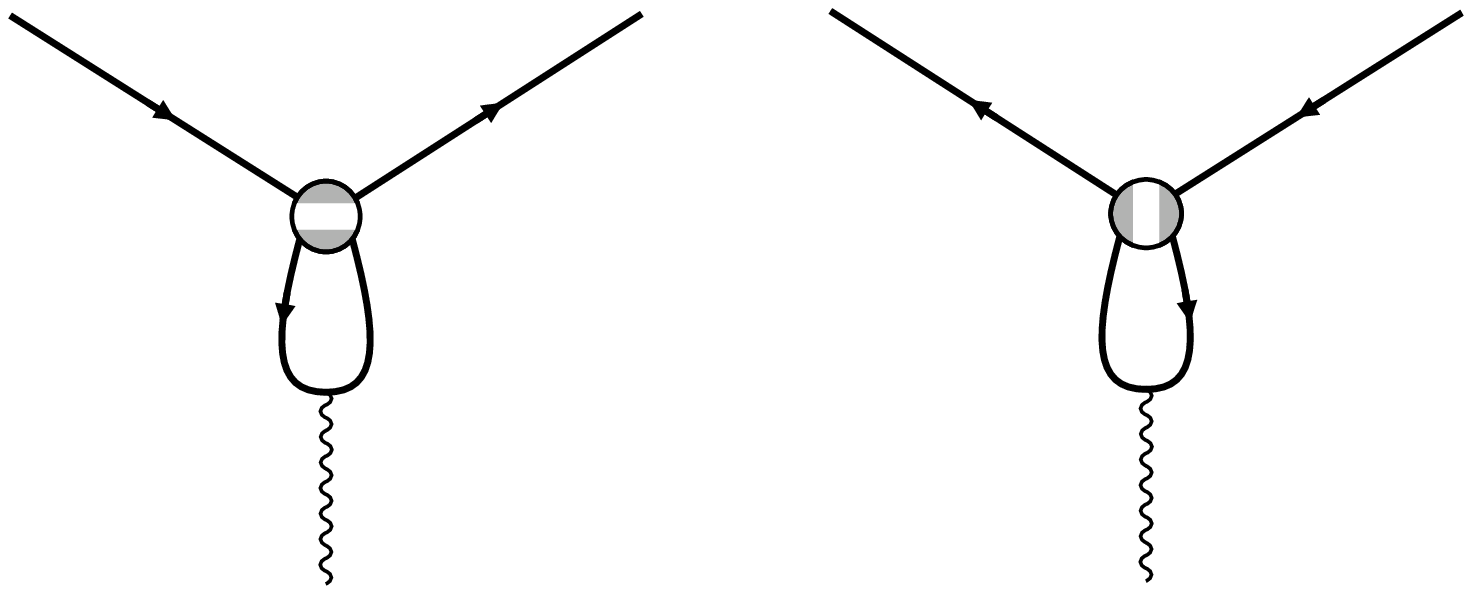}}
\hfil\break
\hbox to 6truein{
\hfil \vtop{\rightskip 0.5truein
\noindent\eightpoint
Figure~2 The penguin diagrams of the walking gauge interactions.
                  \smallskip}\hfil}\hskip 0.5truein
\endinsert\hfuzz 20pt 

\def\marginal{\eta}
For the moment, let us simply ignore this issue, and imagine
that we can find a theory with $r\simeq 2$, or something
equivalent to it.  Take the coupling $\alpha\lappeq\alpha_c$.
The leading four-fermion operator will then have an effective
scaling dimension $4+\marginal$, $\marginal\ll 1$.  I will attempt
to mimic the effect of the large anomalous dimension in
diagrams by simply multiplying any loop containing an $\Op-$
insertion by $(k/\Lambda)^{\marginal-2}$, where $k$ is the loop
momentum.  Now, while $\Op-$ acquires a large negative
anomalous dimension, $\Op+$ acquires a large positive
anomalous dimension.  As a result, it will disappear from
the theory even more quickly than in the weak-coupling
limit of the walking gauge interactions, and we can ignore it.
What about penguins?
If we examine the diagrams of fig.~2, then the loop is naively
quadratically divergent.  Gauge invariance (in the form of derivatives
in the penguin) reduces this to a logarithmic divergence in
the weak-coupling case, so that the diagram indeed generates
a mixing of the flavor-singlet piece of $\Op-$ with the
WGI penguin.  In the assumed strong-coupling scenario, however,
there are nearly two additional powers of the loop momentum, so
that the diagram becomes convergent, no mixing is generated,
and we can safely ignore the penguin.

Note also that while certain operators with more fermion fields also
may acquire negative large anomalous dimensions, their over-all
scaling dimension will presumably remain larger than four, and
hence it is consistent for them to be small.  For example,
the third power of the fermion mass operator, $(\Wfbar\Wf)^3$, might well
acquire an anomalous dimension of $-3$ as $\alpha\rightarrow\alpha_c$;
but this will still leave it an operator of dimension six, and
hence irrelevant.

For nearly marginal operators, there are in
principle important logarithmic self-renormalization effects.
To compute the beta function for the four-fermion coupling $f_{-}$
associated with $\Op-$, we must consider the graphs depicted in
figs.~3.
We can put in an explicit cutoff $p\ll\mu\ll\Lambda$, and differentiate
with respect to $\ln\mu$; this leads to the following expressions,
$$\eqalign{
\beta_{-} &= {l+1\over 2 l}\LB 3 l+7 - N_w (l-3)\RB\;,\cr
\beta(f_{-}) &= \marginal f_{-} - {\beta_{-}\over 32\pi^2}f_{-}^2\;. \cr
}\eqn\betaF$$
(Given the treatment of the anomalous dimension here,
the wavefunction renormalization graphs of fig.~4 have the wrong Lorentz
structure to contribute, and vanish in the presence of a cutoff.)
In computing
the second term, I have set $\eta=0$ since it leads only to terms
of $\Ord(f_{-}^2\eta)$, which is effectively higher order.
I have also ignored any mixing induced by these diagrams with
$\Op+$, since the latter has a very large effective scaling dimension
due to the walking gauge interactions.
Since the operator $\Op-$ started out only nearly marginal rather than
exactly marginal, the beta function starts out at order $f_{-}$ rather
than $f_{-}^2$.  The first term, as expected, has the same sign as the
coupling itself, and so favors an infrared-free coupling.  The
second term, however, is strictly negative for appropriate
choices of $l$ and $N_w$; if $f > f_c = 32\pi^2\eta/\beta_{-} $
it will overcome the leading term, and the beta function will
be negative.  For an attractive coupling, $f>0$, this implies
asymptotic weakness, or more importantly for us, a large value
in the infrared.

\topinsert\hfuzz 40pt
\noindent\hbox to 6truein{
\epsfxsize=6.5truein
\noindent\epsfbox{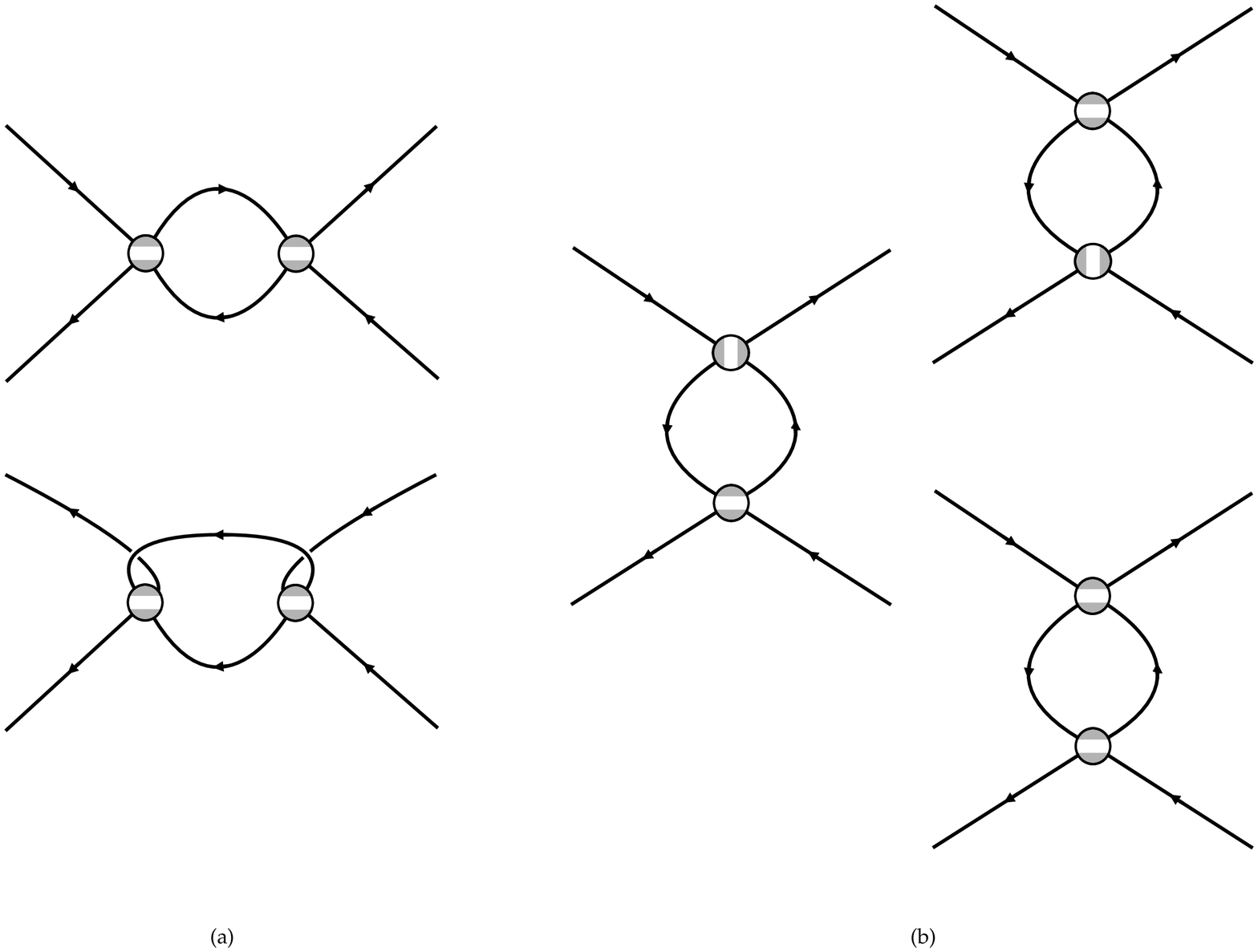}}
\hfil\break
\hbox to 6.5truein{
\hfil \vtop{
\noindent\eightpoint
Figure~3 Contributions to the self-renormalization of the four-fermion
operator.
                  \smallskip}\hfil}
\endinsert\hfuzz 20pt

\topinsert\hfuzz 40pt
\noindent\hbox to 6truein{
\epsfxsize=5.5truein
\noindent\hskip0.5truein\epsfbox{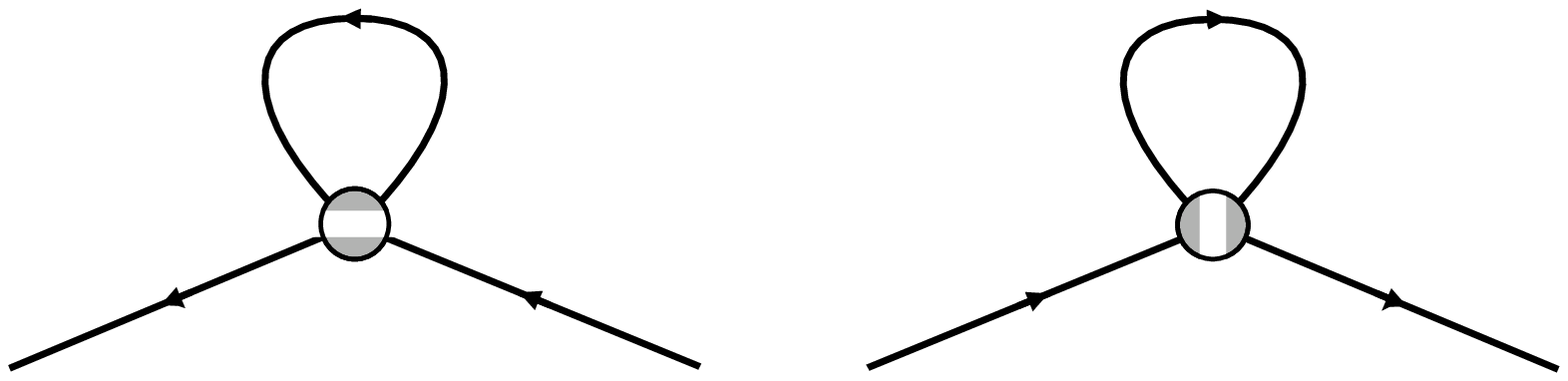}}
\hfil\break
\hbox to 6truein{
\hfil \vtop{\rightskip 0.5truein
\noindent\eightpoint
Figure~4 Contributions of the four-fermion operator to the fermion wavefunction
renormalization.
                  \smallskip}\hfil}
\endinsert\hfuzz 20pt

Within the rainbow approximation considered here, the value of the
coupling $f_{-}$ has no effect on the chiral-symmetry breaking
scale; so if the WGI coupling walks slowly enough, $f_{-}$ can
grow to be quite large down at the scale $\XSB$ where $\alpha$ finally exceeds
$\alpha_c$, and the chirally-symmetric vacuum becomes unstable.
At this scale, the WGI binds the fermions into Goldstone bosons,
and a spectrum of mesons.  What is the effect of $\Op-$ on
the masses of the vector mesons?  Now, it is a composite
operator, and it does not make sense to treat it as a product
of currents; but if we fierz the terms into
products of WGI-singlets,
$$
{f_{-}(\XSB)\over \XSB^2} \L{\XSB\over\Lambda}\R^{\marginal}
 \LB {l+1\over l} \Wfbar T^a\gamma_\mu P_L\Wf\,\Wfbar T^a\gamma^\mu P_L\Wf
     - {l^2-1\over 2 l^2}
        \Wfbar \gamma_\mu P_L\Wf\,\Wfbar \gamma^\mu P_L\Wf\RB\;,
\eqn\MassShift$$
and indulge in the treif temptation to split the operator anyway,
then we will see by analogy with equation~(\BasicL) that the
attractive interaction in the $SU(l)$-adjoint channel
gives the corresponding vector mesons a large negative contribution
to their mass, while the repulsive interaction in the $SU(l)$-singlet
channel presumably gives the singlet meson a positive contribution
to its mass.  The constituent mass of the $\Wf$ plays the role of $M$
in the original Lagrangian~(\BasicL).
Of course, the axial-vector state
receives other contributions to its mass from physics associated
with the spontaneous breaking of the axial symmetry; but the
pure vector state should be unaffected, and thus light.  The
(global) vector symmetry remains exact ignoring weaker interactions,
and can play the role of a custodial symmetry.

The Lagrangians of
BHL-type models also respect chiral symmetry, but their
four-fermion operators are precisely the $R\otimes \overline{R}$
(or equivalently left-right mixing) operators we sought to
avoid earlier.  If we add such operators to a walking gauge theory,
it is plausible that their main effects will be in scalar rather
than vector channels.

The models with vector-channel four-fermion operators
have three different energy regimes: (i) the region stretching
from the high-energy cut-off $\Cutoff$, down to a scale $\Strong$ where
the four-fermion operators become strong, in which all interactions other
than the walking gauge interaction are weak and can be discussed sensibly
in perturbation theory (in particular, explicit breaking of the
chiral symmetries of the WGI can be discussed perturbatively), and in
which some of the four-fermion operators have negative beta functions
and thus grow as we scale down to lower energies;
(ii) the region stretching from the scale $\Strong$
(where some of the four-fermion couplings become appreciable)
 down to a
chiral symmetry-breaking or confinement scale $\XSB$, in which several
interactions are strong and only symmetry arguments will
really tell us anything;
(iii) the region below $\XSB$, in which we will have a chiral effective
Lagrangian with its derivative expansion, along with a light vector
field with (weak) gauge-like couplings, which can again be treated
in perturbation theory.

To introduce ordinary fermions into the theory, we may add terms
of the following form to $\Lag$,
$$
{1\over\Lambda^2}
\Wfbar T^a\gamma_\mu P_L \Wf\,\psibar_L  T^a\gamma_\mu\psi_L\;,
\anoneqn$$
where $\psi$ is an ordinary light fermion.
These will eventually yield couplings of the composite vector
meson to the light fermion.  (One presumably must ensure that
these couplings are non-anomalous, else the light fermion loops
will give an additional contribution of order the chiral-symmetry
breaking scale times powers of the coupling to the vector mass.)

An unusual feature of these models in the first, high-energy, region
is the possibility of having terms in the Lagrangian which violate
the $SU(l)$ symmetry {\it explicitly\/}, so long as they
are sufficiently small (in which case they should not disturb the scenario
explained earlier).  Among such terms we can
consider terms which will give rise to light fermion masses,
$$
\Wfbar \Wf\,\psibar M^l \psi\;.
\anoneqn$$
Indeed, if in building extensions of the Standard Model
we follow the CTSM
philosophy~[\ref\CTSM{R. S. Chivukula and H. Georgi,
\PLB 188:99 (1987); \PRD 36:2102 (1987)}] of assuming that the only
flavor-violating quantities in the high-energy theory are proportional
to the up- and down-quark mass matrices, then all FCNCs in these
models will be suppressed by the usual GIM
mechanism~[\ref\GIM{S. L. Glashow, J. Iliopoulos, and L. Maiani,
 \PRD 2:1285 (1970)}].

I now return to the issue of the $r$ coefficient.  In general,
the coupling $f_{-}$ at the original cutoff scale $\Cutoff$ will
have some finite value.  Even if the anomalous dimension generated
by the WGI alone is not sufficiently large to make the four-fermion
operator $\Op-$ nearly marginal, the combination of the WGI
and the four-fermion coupling itself might suffice.  One might
explore this possibility by examining the combined
Schwinger-Dyson equations, but I will not do so here.

There is another possibility which is more attractive.
We may note that none of the above discussion depended in any
essential way on the eventual breaking of the chiral symmetry; the latter
only served to define the scale at which the fermions
become became bound into light vectors.
This suggests that the same mechanism should also work in the
rather different class of chiral gauge theories.

Consider, for example, $SU(N)$ theories with
left-handed Weyl fermion content of $l$ conjugates of
the symmetric tensor $S^{xy}$ and
$l(N+4)$ fundamentals $\Wf_{z}$.  These theories contain an
$SU(l)\times SU(l[N+4])\times U(1)$ global symmetry, with
the symmetric tensors carrying a charge of $-1$ and the fundamentals
a charge of $(N+2)/(N+4)$.
In discussing
these models, I shall make two assumptions.  The first is that
the global $U(1)$ is not spontaneously broken; this is in fact
the case at at large $N$ [\ref\largeN{E. Eichten, R. D. Peccei,
J. Preskill, and D. Zeppenfeld, \NPB268:161 (1986)}].
The second is that
there is no critical coupling per se, but only a cross-over in description
of the theory from the fundamental fields to composite ones at a scale
of order the confinement scale
This seems reasonable in view of the fact that, given the
first assumption, there is no
Lorentz- and
gauge-invariant order parameter distinguishing the high-energy
from the low-energy theory.  It may make sense to apply similar assumptions
to some of the so-called moose
models~[\ref\Moose{H. Georgi, \NPB 266:274 (1986)}].

In this case, it appears sensible to define instead
the `critical coupling' $\alpha_c'$ as the
value of the gauge coupling $\alpha$ at which the leading
chirally-invariant four-fermion
operator becomes exactly marginal.  For general $N$, this operator is
$$
O_{a} = {1\over\Lambda^2}
  \overline{S}_{xy}T^a\gamma_\mu S_L^{x\hat y}\,
   \Wfbar^y T^a\gamma^\mu\Wf_{L\hat y}\;,
\eqn\chiralFourFermi$$
where $T^a$ are the generators
of $SU(l)$ or one of its subgroups, possibly in a direct sum of copies
of the fundamental representation when acting on the $\Wf$ fields.
In this theory, there simply is no fermion bilinear, and so we need not
worry about the $r$ coefficient.
The beta function for this coupling is a bit different from that
given in equation~(\use\betaF), as only the diagrams of fig.~3~(a)
will contribute, and the analog
of $\beta_{-} = (3 l^2 + 8 N^2 + 2 + N (14+l^2))/(l (N+1))$.
  (This operator will mix with current-current
type operators containing only symmetric tensors or only fundamental
fermions; but as these have smaller WGI-generated anomalous
dimensions in the rainbow approximation and are thus plausibly
not nearly marginal, I will ignore them.  It will also mix
with an operator with the same fields and color structures as
that of equation~(\use\chiralFourFermi), but with the identity
replacing the flavor matrices.  This operator $O_1$ has the same
WGI-generated anomalous dimension
as $O_{a}$, and thus will be generated
by renormalization-group flow.  It turns out, however, that
the relevant beta-coefficient matrix has only one positive eigenvalue;
the corresponding infrared-attractive eigenvector is predominantly
$O_a$, with a small repulsive
admixture of $O_1$, which leaves the qualitative conclusions of
equation~(\use\MassShift) intact.)

Given the assumption that the $U(1)$ global symmetry is unbroken,
there will necessarily be massless fermions in the theory.  The most
likely possibility is the straightforward generalization of the
$l=1$ model, in which the light fermions transform as an
antisymmetric tensor under a surviving $SU(N+4)$ symmetry.  In the
case considered here, we will presumably have $l$ copies of such
an antisymmetric tensor.  If $l=2$, it is possible, and
in the presence of a strong four-fermion operator of the
form in equation~(\use\chiralFourFermi) even plausible, for an $SU(2)$
global symmetry to be preserved.  (For larger $l$ it does
not seem possible for the $SU(l)$ to be preserved; presumably
some maximal vector-like subgroup will be.  This subgroup
may well depend on the four-fermion operators present in the
theory at the confinement scale.).

As an example of a walking chiral gauge theory, we can take
a model of this class with $N=4$, and $l=2$.  In this case,
if we assume that the critical coupling $\alpha_c'$ is given
by $\gamma_0 = -1$, we find
$$
\alpha_c' = 0.47,\hskip 1cm \beta(\alpha_c')/\alpha_c' = -0.15\;,
\anoneqn
$$
with the two-loop formula.  (If we prefer to use the three-loop
{MS}-scheme formula [\ref\ThreeLoop{O. V. Tarasov, A. A. Vladimirov,
and A. Yu. Zharkov, \PLB 93:429 (1980)}], we can take $N=7$ and
$l=2$ as an example, for which $\alpha_c' = 0.27$ and
$\beta(\alpha_c')/\alpha_c' = -0.24$.)

The general outline of the discussion in vector-like models
continues to apply,
 so long as we assume that a large value for the
four-fermion coupling does not trigger `premature
confinement': then the value of the four-fermion coupling at the
confinement scale
where we trade the high-energy description for the low-energy
one will be large, and correspondingly the vector meson will be light
compared to the confinement scale.

In this particular $SU(4)$ model, we will also have $28$ fermions transforming
as doublets under the $SU(2)$ to which the light vector meson couples,
and as an antisymmetric tensor under the suriving $SU(8)$ global symmetry.
This is an attractive feature of these models: they naturally
generate light fermions (massless in the absence of explicit breaking of the
$U(1)$) which transform non-trivially under the effective gauge
symmetry.
For example, the left-handed fermions of the Standard Model might
be composites, while the right-handed ones remain fundamental; mass
terms could emerge from four-fermion terms in the high-energy
theory which violate the $SU(2)$ symmetry explicitly.
An amusing variant of Kaplan's mechanism~[\ref\Kaplan{D. B. Kaplan,
\NPB 365:259 (1991)}] in technicolor
theories, in which ordinary fermions acquire
their masses through mixing with technifermions via higher-dimension
operators, may be possible, in which it is not the different
naive dimensions of these operators that leads to a hierarchy
but rather the different WGI-induced anomalous dimensions.

More generally, we
can also consider models which have both chiral and vector-like
symmetries, for example:
$$\eqalign{
SU(N)_w:&\hskip 1cm l \times S \oplus [l (N+4) + v] \times F
\oplus v\times \overline{F}\;;\cr
&{\rm or}\cr
SU(N)_w:&\hskip 1cm l \times A \oplus [l (N-4) + v] \times F
\oplus v\times \overline{F}\;.\cr
}\anoneqn$$
Here, $S$ and $A$ are the conjugates of the symmetric and antisymmetric
tensor representations, respectively.
In these models, there is again a chiral-symmetry breaking scale
associated with the vector-like piece, and thus
an $r$ coefficient, but it can
be near (or even larger than) $2$; indeed, for models with
a symmetric tensor,
$$
r = {2 (N+2)\over N+1}
\anoneqn$$
for the four-fermion operators of the kind given in
equation~(\use\chiralFourFermi),
while for those with an antisymmetric tensor,
$$
r = {2 (N-2)\over N-1}
\anoneqn$$
for similar four-fermion operators with the symmetric tensor replaced
by an antisymmetric one (and with $N>6$; for $N=5$ the leading
operator involves only antisymmetric tensors while for $N=6$ there
are two leading operators, one of each kind, which presumably mix).
In the large-$N$ limit, $r=2$ and both $\alpha_c$ and
$\alpha_c'$ are $2\pi/(3 N)$.  Furthermore, if we take $l=2$
and $v= (1+\epsilon) N + O(1)$, we find
$$\beta(\alpha_c)/\alpha_c =
-{11\over 27} + {25\over 54}\epsilon
\anoneqn$$
if we use the two-loop formula,
$$\beta(\alpha_c)/\alpha_c =
-{473\over 2916} + {463\over 729}\epsilon - {14\over 729}\epsilon^2
\anoneqn$$
if we use the three-loop MS one.
In addition to the possibility of explicit breaking of the eventual
gauge-like interaction, one should also keep in mind that different
members of an electroweak doublet do not have to have a common
origin at high energy; for example,
in such models with only partially-chiral representations, it may
be possible for the top quark to be a WGI-baryon, while the
bottom quark remains a light composite.  (This would violate the
custodial $SU(2)$, of course, but we already know that the top quark
mass is large. Whether this is consistent with a light $W$ depends
on the dynamical competition between the attractive four-fermion
interactions and the WGI.)

Finally, I will mention a
more extreme yet richer scenario
that may be possible in theories in which the
running of the four-fermion coupling at large coupling is
itself dependent on the matter content of the theory (as would
be the case, for example, for a variety of moose models).  For
a suitable choice of matter representations, it
too may be able to `walk'.  In such a case
(assuming again that there is no `premature confinement'), there is
plausibly a critical value for the four-fermion coupling at which
operators with yet larger numbers of fermion fields
become marginal, and their
dynamical evolution must be considered as well.

I thank A.~Manohar for discussions on various aspects of walking
gauge theories and operator renormalization.

\vfill\eject\immediate\closeout\rfile
\centerline{{\bf References}}\bigskip\frenchspacing%
\input refs.tmp\vfill\eject\nonfrenchspacing

\bye